# CROSS-DOMAIN SINGLE-CHANNEL SPEECH ENHANCEMENT MODEL WITH BI-PROJECTION FUSION MODULE FOR NOISE-ROBUST ASR

*Fu-An Chao[1], Jeih-weih Hung[2], Berlin Chen[1]*

[1]National Taiwan Normal University, Taipei, Taiwan
[2]National Chi Nan University, Nantou, Taiwan

{fuann, berlin}@ntnu.edu.tw, jwhung@ncnu.edu.tw

**ABSTRACT**

In recent decades, many studies have suggested that phase information is crucial for speech enhancement (SE), and time-domain single-channel speech enhancement techniques have shown promise in noise suppression and robust automatic speech recognition (ASR). This paper presents a continuation of the above lines of research and explores two effective SE methods that consider phase information in time domain and frequency domain of speech signals, respectively. Going one step further, we put forward a novel cross-domain speech enhancement model and a bi-projection fusion (BPF) mechanism for noise-robust ASR. To evaluate the effectiveness of our proposed method, we conduct an extensive set of experiments on the publicly-available Aishell-1 Mandarin benchmark speech corpus. The evaluation results confirm the superiority of our proposed method in relation to a few current top-of-the-line time-domain and frequency-domain SE methods in both enhancement and ASR evaluation metrics for the test set of scenarios contaminated with seen and unseen noise, respectively.

## 1. INTRODUCTION

In recent years, due to the unprecedented breakthroughs in deep learning and deep neural networks (DNN), the state-of-the-art automatic speech recognition (ASR) systems have reached human parity and exhibited remarkable performance on many tasks as long as the input speech is recorded in a noise-free condition. However, when a well-trained ASR system is deployed into a real environment, the performance might be seriously degraded because of environmental interferences such as noise, reverberation, background speakers, and the like. To mitigate these deteriorating effects so as to improve the real-world ASR performance, researchers and practitioners have developed a great number of techniques. Among these techniques, speech enhancement (SE) has seen widespread adoption as a pre-processing stage to be a crucial component before acoustic modeling in ASR to alleviate environmental interferences.

As for the state-of-the-art SE methods, multi-channel approaches equipped with a microphone array structure usually can behave very well. Comparatively, in most real-world scenarios that lack multiple array microphones, single-channel SE techniques are typically less effective and only yield moderate improvements in speech quality and intelligibility metrics like perceptual evaluation of speech quality (PESQ) [1] and short-time objective intelligibility (STOI) [2]. However, these improved results do not necessarily translate well into the back-end ASR performance. This may be due to the artifacts introduced by the front-end SE module and the discrepancy of the training objectives between SE and ASR. To tackle this problem, several efforts have been made to develop robust single-channel SE methods that could benefit the ASR performance [3]. For example, the joint training scheme [4] was established along this direction.

In particular, time-domain speech signal processing has attracted much interest from both academic and commercial sectors, and persistent research has been conducted on analyzing the time domain of speech signals for noise reduction [5] and speech separation [6]. For instance, convolutional time-domain audio separation network (Conv-TasNet) [6] exhibits outstanding performance on the speech separation task and surpasses many frequency-domain approaches. Compared to frequency-domain approaches, Conv-TasNet implicitly encodes phase information of an input signal to a latent representation with a convolutional operation, which is regulated by its kernel sizes and strides, similar to the short-time Fourier transform (STFT). The underlying notion is also adopted in the design of the single-channel SE module as the front-end of an ASR system [5], and it achieves a substantial improvement even when the ASR system has already been equipped with sophisticated acoustic models (AM) with a multi-condition training (MCT) setting. However, since Conv-TasNet is typically built with a pure DNN-based encoder/decoder architecture, it might fail to perform well when only a small-scale dataset is made available for training its component models [7], probably leading to incorrect estimation of the distribution for speech features.

Building on these observations, in this study we explore to leverage two domains of speech features derived from a 1-D convolution layer and the conventional STFT, respectively, and in turn present a bi-projection fusion (BPF) mechanism to formulate a novel SE framework. This framework exploits cross-domain speech features augmented with their BPF counterpart in order to reduce the undesirable noise effect on speech quality and recognition accuracy. Experimental results show that our proposed method surpasses some other state-of-the-art methods, including the time-domain method Conv-TasNet and the frequency-domain method STFT-TCN [7], in terms of scale-invariant signal-to-noise ratio (SI-SNR) and character error rate (CER).

## 2. RELATED WORK

### 2.1. Speech enhancement algorithms

Considering a discrete-time noisy signal $y[n]$ captured by a single-channel microphone, we can formulate the following equation:

$$y[n] = h[n] * x[n] + d[n], \qquad (1)$$

where $x[n]$ is the target noise-free speech signal, $h[n]$ is the convolutional noise, "$*$" denotes the convolution operation, $d[n]$ is the additive noise, and $n$ is the time index. In this study, we focus exclusively on alleviating the additive noise $d[n]$ from the noisy signal $y[n]$ to recover the speech signal $x[n]$, while convolutional noise $h[n]$ like reverberation or channel mismatch is assumed to be negligible and thus excluded from consideration.

Most of the early efforts on noise reduction analyze noisy speech signals in the acoustic frequency domain of signals via short-time Fourier transform (STFT). Such techniques like Wiener filtering [8], low-rank representation (LRR) [9] employ the statistics drawn from the short-time spectra of speech to reduce noise. Notably, in recent years, the prevalent deep-neural-network (DNN) techniques have been adopted to develop SE algorithms with impressive success. These DNN-wise SE methods can be further classified to be either mapping-based or masking-based [10][11]. For example, Lu *et al.* [10] is the first to employ a deep denoising autoencoder (DDAE) to map the power spectrum of noisy speech to that of its clean counterpart directly. Wang *et al.* [11] proposed to use a deep neural network (DNN) to implicitly predict a time-frequency (T-F) mask so as to be applied to a noisy spectrogram for SE.

It is noteworthy that pursuing an effective mask for speech feature representations has become one of the most predominant directions for the SE research. The initial DNN-based masking algorithms, including, but is not limited to, ideal binary mask (IBM) and ideal ratio mask (IRM) [11], are mostly conducted on the magnitude spectrogram of speech. Recently, many studies have shown that the phase information is crucial to the success of SE [12][13]. To take the phase into account, we can either extract the magnitude and phase parts concurrently in the frequency domain of speech signals, or process speech signals in time domain directly. Both directions have been proven effective and superior to previous attempts that operated only on magnitude spectrograms. As an illustration, [13] proposed phase-aware SE in frequency domain using a two-stream architecture to predict a complex-ratio mask (cRM) and jointly reconstruct the magnitude and phase parts. [5] adopted Conv-TasNet to perform masking-based SE directly in time domain and gains improvements in terms of SE metrics and achieved lower word error rates on an ASR task.

In the following two sub-sections, we briefly introduce two SE processes, which will serve as the component modules of our SE framework to be presented in the next section.

### 2.2. Encoder-decoder architecture

A neural SE algorithm often adopts an encoder-decoder architecture to process the input signal and in turn output the enhanced signal as in [6]. First, we divide the input speech signal into overlapping segments of length $L$, represented by $x_k \in \mathbb{R}^{L \times 1}$, where $k$ denotes the segment index out of a total of $K$ segments. As such, by concatenating all segments horizontally, we can obtain an input matrix $\mathbf{X} \in \mathbb{R}^{L \times K}$ whose $k^{\text{th}}$ column is $x_k$.

The encoder is then designed to transform $\mathbf{X}$ into a $N$-dimensional representation $\mathbf{W} \in \mathbb{R}^{N \times K}$, which can act as either a conventional STFT operated in the frequency domain or a trainable linear transformation such as a 1-D convolution operation learned through the network in the time domain. The encoding step can be formulated as an operation as follows:

$$\mathbf{W} = \mathcal{F}(\mathbf{UX}), \qquad (2)$$

where $\mathbf{U} \in \mathbb{R}^{N \times L}$ contains $N$ basis functions, and $\mathcal{F}(\cdot)$ is an optional activation function, which can be the rectified linear unit (ReLU) in the time-domain representation [6] to ensure non-negative outputs.

The decoder is used to reconstruct the waveform from the encoded representation $\mathbf{W}$ (or its enhanced version), and it can also be formulated as another matrix multiplication:

$$\hat{\mathbf{S}} = \mathbf{VW}, \qquad (3)$$

where $\hat{\mathbf{S}} \in \mathbb{R}^{L \times K}$ is the reconstructed version of input matrix $\mathbf{X}$, and $\mathbf{V} \in \mathbb{R}^{L \times N}$ consists of the basis functions involved in the decoder. For SE conducted in the frequency domain, the decoder is usually the inverse short-time Fourier transform (ISTFT), while it can be a 1-D transpose convolution operation in the time domain. The finally enhanced waveform is generated from the reconstructed segments in $\hat{\mathbf{S}}$ using the overlap-add method.

### 2.3. Mask estimation network

As for masking-based SE, it is typical to employ a DNN model to estimate a mask applying to a noisy mixture representation $\mathbf{W}$. To effectively capture the temporal information and consider the long-term dependency of frames in the input signal, the associated DNN can be created by stacking BLSTM layers or dilated convolution layers. The output of the mask estimation network can directly be a single mask for the target speech, and the other choice of this network is to predict two masks respectively for speech and noise, denoted by:

$$[\mathbf{M}_x, \mathbf{M}_d] = \mathcal{M}_\theta(\mathbf{W}), \qquad (4)$$

where $\mathcal{M}_\theta(\cdot)$ represents the mask estimation network, and $\mathbf{M}_x$ and $\mathbf{M}_d$ are the masks for target speech and noise, respectively.

After that, we can obtain the enhanced representation $\mathbf{D}_x$ by applying the speech mask $\mathbf{M}_x$ to the mixture representation $\mathbf{W}$:

$$\mathbf{D}_x = \mathbf{M}_x \odot \mathbf{W}, \qquad (5)$$

where $\odot$ is the element-wise multiplication. The finally enhanced waveform can be created or decoded from $\mathbf{D}_x$ as Eq. (3).

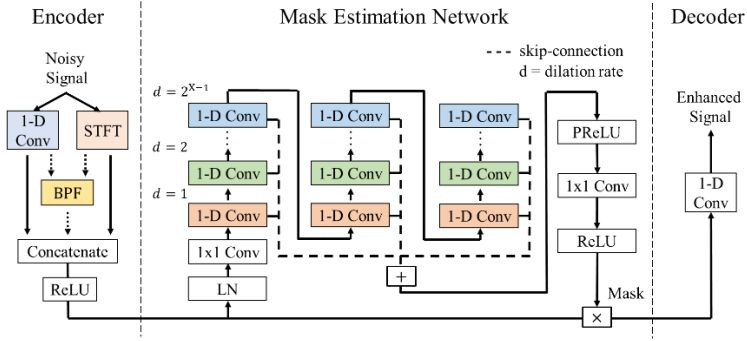

**Fig. 1**: The schematic diagram of the CD-TCN system.

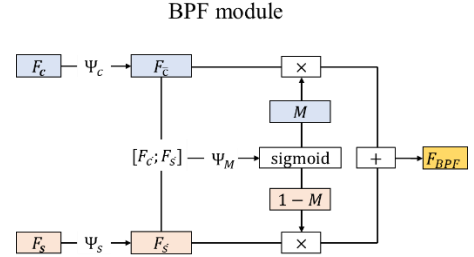

**Fig. 2**: Proposed BPF-module.

## 3. PROPOSED METHOD

In this section, we introduce a novel single-channel SE framework, which consists of a cross-domain encoder, a mask estimation network and a decoder. This novel SE framework can harness the synergistic advantages of the two SE structures elucidated in Sections 2.2 and 2.3, and we refer it to as Cross-Domain Temporal Convolutional Network, with a short-hand notation "CD-TCN" hereafter. The block diagram of CD-TCN is depicted in Figure 1, where a network structure based on the temporal convolution network (TCN) is used to implement mask estimation. In the following, we describe the ingredient components of CD-TCN.

### 3.1. Cross-domain encoder

As shown in Figure 1, the input noisy speech signal is converted to time-domain and frequency-domain feature representations, respectively (*cf.* Section 2.2), and both of them preserve the phase information of the input signal. The time-domain feature is derived from a trainable 1-D convolution layer, while the frequency-domain feature is generated from the conventional STFT with discrete Fourier basis functions, which are non-trainable. Note here that, to take the phase part into account, the frequency-domain (STFT) features are complex-valued spectrograms, which have real and imaginary parts that are structured as in [7]. These two-domain feature representations are calculated concurrently, and then they are spliced together to perform the subsequent mask estimation.

### 3.2. Bi-projection fusion module

To further extract the information shared across the features from the two different domains, we seek to develop a fusion scheme which is expected to learn the relation between each other. Conceptually similar to the recent attempts of the computer vision community [14][15], here a novel bi-projection fusion (BPF) module is proposed and operationalized as shown in Figure 2, whose details are as follows: given the two branches of feature representation, namely the time-domain feature $F_c$ and frequency-domain feature $F_s$, we first individually convert them to two new features, $F_{\bar{c}}$ and $F_{\bar{s}}$, both of which have the same dimension. Next, in order to exploit both branches and balance the fusion procedure, we use the concatenation of $F_{\bar{c}}$ and $F_{\bar{s}}$ to estimate a ratio mask $M$. Finally, we apply this mask to both branches and in turn generate the BPF feature. The whole procedure can be formulated as:

$$F_{\bar{c}} = \Psi_c(F_c, \theta_c), \tag{6}$$

$$F_{\bar{s}} = \Psi_s(F_s, \theta_s), \tag{7}$$

$$M = \text{sigmoid}(\Psi_M(\text{concat}(F_{\bar{c}}, F_{\bar{s}}), \theta_M)), \tag{8}$$

$$F_{BPF} = M \odot F_{\bar{c}} + (1 - M) \odot F_{\bar{s}}, \tag{9}$$

where $\Psi_c$, $\Psi_s$, and $\Psi_M$ are projection layer operations with parameters $\theta_c$, $\theta_s$, and $\theta_M$, respectively, $M$ is the estimated ratio mask with the same dimension as $F_{\bar{c}}$ and $F_{\bar{s}}$, and $F_{BPF}$ is the final BPF feature.

Finally, the obtained BPF feature serves as an auxiliary input, which is concatenated with the time-domain and frequency-domain features $F_c$ and $F_s$ and then passed to the mask estimation network, as shown in Fig. 1. To the best of our knowledge, CD-TCN is the first attempt to integrate cross-domain clues of a speech signal, as well as their BPF features, for effective SE. In a move to encourage more instantiations in the future, we plan to provide a publicly available version of the code used for our experiments.

## 4. EXPERIMENTAL SETUP

We conduct an extensive set of empirical experiments on AISHELL-1 [16], which is an open-source Mandarin ASR benchmark corpus containing 400 speakers and over 170 hours of speech data. To evaluate our proposed method, we created some synthetic data from the original training set and designed our test data at different signal-to-noise ratios (SNRs) using a variety of online-available noise databases.

As for the training phase, a total of 2,553 noise recordings were collected from the MUSAN dataset [17], the DEMAND dataset [18], the QUT-NOISE dataset [19], and the Environmental

| AM Model | SE Model | Test sets with seen noises | | | | | |
|---|---|---|---|---|---|---|---|
| | | -5dB | | 5dB | | 15dB | |
| | | CER | SI-SNR | CER | SI-SNR | CER | SI-SNR |
| Baseline | – | 81.41 | -4.94 | 43.81 | 5.02 | 15.14 | 15.02 |
| MCT-AM | – | 58.19 | -4.94 | 18.75 | 5.02 | 8.66 | 15.02 |
| | STFT-TCN | 41.77 | **7.92** | 16.73 | 15.16 | 8.36 | 20.19 |
| | Conv-TasNet | 40.70 | 6.93 | 15.78 | 16.08 | 8.17 | 20.85 |
| | CD-TCN | 40.15 | 7.09 | 14.30 | 16.30 | 8.10 | 20.91 |
| | + BPF | **40.08** | 7.29 | **13.58** | **16.43** | **8.08** | **21.10** |

**Table 1**: SI-SNR (dB) and CER (%) results of different SE systems for the seen-noise test scenario.

| AM Model | SE Model | Test sets with unseen noises | | | | | |
|---|---|---|---|---|---|---|---|
| | | -5dB | | 5dB | | 15dB | |
| | | CER | SI-SNR | CER | SI-SNR | CER | SI-SNR |
| Baseline | – | 88.42 | -4.97 | 50.58 | 5.01 | 16.92 | 15.01 |
| MCT-AM | – | 70.90 | -4.97 | 25.37 | 5.01 | 10.45 | 15.01 |
| | STFT-TCN | 51.69 | **8.01** | 20.40 | 15.12 | 10.03 | 20.43 |
| | Conv-TasNet | 50.12 | 7.58 | 18.74 | 15.72 | 9.69 | 21.15 |
| | CD-TCN | 49.21 | 7.74 | 18.51 | 15.74 | 9.62 | 21.17 |
| | + BPF | **49.04** | 7.83 | **17.82** | **15.83** | **9.52** | **21.30** |

**Table 2**: SI-SNR (dB) and CER (%) results of different SE systems for the unseen-noise test scenario.

Background Noise dataset [20][21]. All of the clean speech and noise recordings were resampled to be 16-kHz. Based on these recordings, we created a multi-condition training (MCT) data set (denoted by "MCT data" hereafter) at the SNR of 5 dB, which had the same size as the original training set, i.e., the 120,098 simulated utterances.

Regarding the test phase, the original test set was corrupted by noise recordings from the aforementioned noise databases to create three test sets at three SNRs: -5, 5, and 15 dB, respectively. In addition, we used different types of noise to create another test set which simulated the unseen noisy environments. The noisy data was generated by adding the noise recordings complied from the Nonspeech dataset [22], which contained 100 types of non-stationary noise at -5, 5, and 15 dB of SNRs, respectively.

### 4.1. SE system configuration

In the SE experiments, we select 100 utterances per speaker from the original training set to learn the SE systems, where all data instances are mixed with the MCT noise at the 5dB SNR level. We set the configuration of the mask estimation network to be a temporal convolutional network (TCN) with the hyperparameter $X=8, R=3, B=128, H=512, S=128$ and $P=3$, which follows the settings in [6]. All of the SE systems for comparison here are trained to minimize the negative scale-invariant signal-to-noise ratio (SI-SNR) loss [6], and they output only enhanced speech. Notably, the permutation invariant training (PIT) previously used for the speech separation task in [6] is not employed here.

In particular, we choose two state-of-the-art SE approaches for comparison: STFT-TCN [7] and Conv-TasNet, which perform in the frequency-domain and the time-domain, respectively. In STFT-TCN, we use STFT and inverse STFT (ISTFT) as the encoder-decoder architecture. The FFT size is set to 512, and the frame size and frame shift are set to 64 and 32, respectively, with a Hanning window function. In Conv-TasNet, we use an architecture similar to those presented in [5] and add skip-connections in the used TCN, which can result in better SE performance. The number of filters is set to 512, and the window size is set to 16 with 1/2 overlapping.

As for the newly presented CD-TCN, we take both of the time-domain and frequency-domain features with 256 dimensions, and the window size is set to 16 with 1/2 overlapping. The number of hidden units in the projection layer of the BPF module is set to 128 to obtain 128-dimensional BPF features.

### 4.2. ASR system configuration

For the back-end ASR system, we use Kaldi [23] to build hybrid DNN-HMM acoustic models following the standard training recipe provided by [16]. Note that, the DNN model is replaced by a factorized time-delay neural network (TDNN-F) and trained with the lattice-free MMI (LF-MMI) objective function, which is shown to achieve better results on the original test set.

In addition to the baseline system, we make use of the aforementioned MCT data in combination with the original raw data to perform multi-condition training (MCT). The resulting

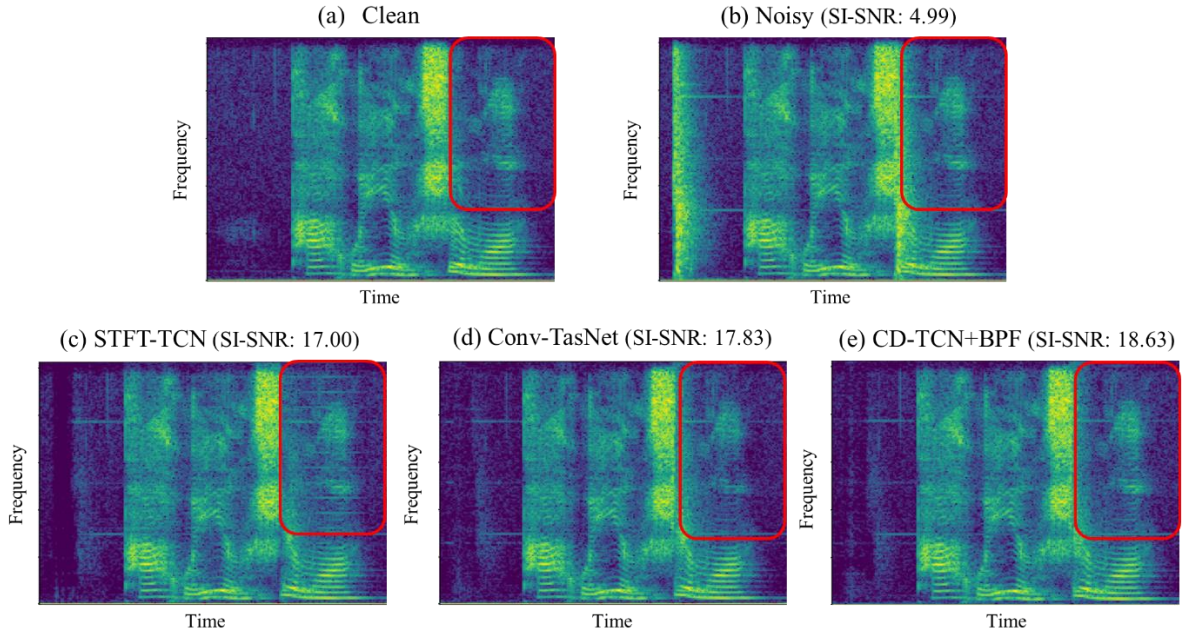

**Fig. 3**: The enhanced spectrograms of a speech utterance corrupted by clanking noise at 5 dB.

acoustic models are denoted by MCT-AM in the following discussions.

## 5. EXPERIMENTAL RESULTS AND DISCUSSIONS

In this section, we use the scale-invariant signal-to-noise ratio (SI-SNR) and character error rate (CER) as the metrics to evaluate the aforementioned different front-end SE systems, which are tested with respect to two distinct scenarios: seen-noise and unseen-noise. The results are shown in Tables 1 and 2, respectively.

### 5.1. Effect of different SE systems

Looking into the results on the seen-noise case shown in Table 1, we make three noteworthy observations. First, compared with the baseline where only the original training set is used for creating acoustic models (AM), the further adoption of MCT data enables the resulting MCT-AM to gain a significant reduction in CER.

Second, all of the SE front-ends compared here can promote the SI-SNR scores at three SNR cases. The STFT-TCN method behaves the best for the lowest SNR case (-5 dB), while the presented CD-TCN method plus BPF outperforms the others at the moderate and high SNRs (5 dB and 15 dB).

Third, surprisingly, with MCT-AM as the back-end acoustic models, all of the SE methods can achieve not only higher SI-SNR scores but also improve ASR performance for all SNR conditions, especially in the extremely noisy scenario (-5 dB). In particular, CD-TCN (without BPF) delivers lower CERs than the state-of-the-art Conv-TasNet and STFT-TCN methods, and CD-TCN that additionally makes use of BPF features can lead to further CER reduction. The improved SI-SNR and CER results, therefore, clearly support our hypothesis that the BPF feature, which integrates the time-domain and frequency-domain features, benefits the subsequent mask estimation and further mitigates the noise effect.

### 5.2. Generalization to unseen-noise test scenario

Next, we examine the generalization capability of the presented method for SE by observing the evaluation results shown in Table 2, where the noise sources in the test set are unseen during the AM training process.

As we can see, the results in Table 2 show a close tendency with those in Table 1. The presented CD-TCN plus BPF method brings the most significant SI-SNR improvement for almost all SNR cases and obtains the lowest CER for all conditions, even though the test utterances contain unseen noise sources. Consequently, we further confirm that CD-TCN plus BPF can show excellent performance in promoting the quality and recognition accuracy of speech signals under mismatched noise environments.

### 5.3. Visualization of spectrograms

To provide a visual inspection on the effectiveness of our method on SE, we plot the spectrograms of one noisy speech utterance at 5 dB SNR level, as well as its clean and enhanced counterparts by using different SE front-ends. As shown in Figure 3, each SE system markedly suppresses the interrupted noise, but still with some residuals. Furthermore, as the area outlined in the red box, STFT-TCN introduces some artifacts on high-frequency bands that might result in a poor enhancement outcome. Comparatively, the proposed CD-TCN with BPF performs better than STFT-TCN and Conv-TasNet on the reconstruction of the speech signal and thus leads to the best results (the highest SI-SNR scores).

## 6. CONCLUSION

In this study, we have proposed a cross-domain speech enhancement framework for noise-robust ASR and introduced a bi-projection fusion (BPF) module to leverage both time-domain and frequency-domain features. This novel framework has shown to considerably improve speech quality (SI-SNR) and reduce the recognition error (CER) simultaneously. Compared with the state-of-the-art SE methods, STFT-TCN and Conv-TasNet, the presented CD-TCN plus BPF achieves even better performance in speech enhancement and noise-robust speech recognition. Therefore, we believe that CD-TCN plus BPF offers a promising avenue for future SE and ASR developments.

## 7. ACKNOWLEDGEMENT

This research is supported by the Ministry of Science and Technology (MOST), Taiwan, under Grant Number MOST 109-2634-F-008-006- through Pervasive Artificial Intelligence Research (PAIR) Labs, Taiwan, and Grant Numbers MOST 108-2221-E-003-005-MY3 and MOST 109-2221-E-003-020-MY3. Any findings and implications in the paper do not necessarily reflect those of the sponsors.

## REFERENCES

[1] A. W. Rix, J. G. Beerends, M. P. Hollier, and A. P. Hekstra, "Perceptual evaluation of speech quality (PESQ) - a new method for speech quality assessment of telephone networks and codecs," in *Proc. ICASSP*, pp. 749–752, 2001.

[2] C. H. Taal, R. C. Hendriks, R. Heusdens, and J. Jensen, "An algorithm for intelligibility prediction of time–frequency weighted noisy speech," *IEEE/ACM Transactions on Audio, Speech, and Language Processing*, vol. 19, no. 7, pp. 2125–2136, 2011.

[3] K. Tan and D. Wang, "Improving robustness of deep learning based monaural speech enhancement against processing artifacts," in *Proc. ICASSP*, 2020.

[4] T. Menne, R. Schlüter and H. Ney, "Investigation into joint optimization of single channel speech enhancement and acoustic modeling for robust ASR," *arXiv preprint arXiv:1904.09049*, 2019.

[5] K. Kinoshita, T. Ochiai, M. Delcroix, and T. Nakatani, "Improving noise robust automatic speech recognition with single-channel time-domain enhancement network," in *Proc. ICASSP*, pp. 7009–7013, 2020.

[6] Y. Luo and N. Mesgarani, "Conv-TasNet: Surpassing ideal time–frequency magnitude masking for speech separation," *IEEE/ACM Transactions on Audio, Speech, and Language Processing*, vol. 27, no. 8, pp. 1256–1266, 2019.

[7] Y. Koyama, T. Vuong, S. Uhlich, and B. Raj, "Exploring the best loss function for DNN-based low-latency speech enhancement with temporal convolutional networks," *arXiv preprint arXiv:2005.11611*, 2020.

[8] N. Wiener, *Extrapolation, interpolation, and smoothing of stationary time series*, New York: WILEY, 1949.

[9] B. Yan, C. Shih, S. Liu, B. Chen, "Exploring low-dimensional structures of modulation spectra for robust speech recognition," in *Proc. INTERSPEECH*, 2017.

[10] X. Lu, Y. Tsao, S. Matsuda, and C. Hori, "Speech enhancement based on deep denoising autoencoder," in *Proc. INTERSPEECH*, pp. 436–440, 2013.

[11] Y. Wang and D. L. Wang, "Towards scaling up classification-based speech separation," *IEEE/ACM Transactions on Audio, Speech, and Language Processing*, vol. 21, pp. 1381–1390, 2013.

[12] K. Paliwal, K. Wojcicki, and B. Shannon, "The importance of phase in speech enhancement," *Speech Communication*, vol. 53, no. 4, pp. 465–494, 2011.

[13] D. Yin, C. Luo, Z. Xiong, and W. Zeng, "PHASEN: A phase-and-harmonics-aware speech enhancement network," *arXiv:1911.04697*, 2019.

[14] F. Wang, Y. Yeh, Min Sun, W. Chiu, and Y. Tsai. "BiFuse: Monocular 360° depth estimation via bi-projection fusion," in *Proc. CVPR*, pp. 462–471, 2020.

[15] Z. Zhang, Z. Cui, C. Xu, Z. Jie, X. Li, and J. Yang. "Joint task-recursive learning for semantic segmentation and depth estimation," in *Proc. ECCV*, 2018.

[16] H. Bu, J. Du, X. Na, B. Wu, and H. Zheng, "AISHELL1: An open-source Mandarin speech corpus and a speech recognition baseline," in *Proc. O-COCOSDA*, pp. 1–5, 2017.

[17] D. Snyder, G. Chen, and D. Povey, "MUSAN: A music, speech, and noise corpus," *arXiv preprint arXiv:1510.08484*, 2015.

[18] J. Thiemann, N. Ito, and E. Vincent, "The diverse environments multichannel acoustic noise database: A database of multichannel environmental noise recordings," *The Journal of the Acoustical Society of America*, vol. 133, no. 5, pp. 3591–3591, 2013.

[19] D. B. Dean, S. Sridharan, R. J. Vogt, and M. W. Mason, "The QUT-NOISE-TIMIT corpus for the evaluation of voice activity detection algorithms," in *Proc. INTERSPEECH*, pp. 3110–3113, 2010.

[20] F. Saki, A. Sehgal, I. Panahi, and N. Kehtarnavaz, "Smart phone-based real-time classification of noise signals using subband features and random forest classifier," in *Proc. ICASSP*, pp. 2204–2208, 2016.

[21] F. Saki and N. Kehtarnavaz, "Automatic switching between noise classification and speech enhancement for hearing aid devices," in *Proc. EMBC*, pp. 736–739, 2016.

[22] G. Hu and D. Wang, "A tandem algorithm for pitch estimation and voiced speech segregation," *IEEE Transactions on Audio, Speech, and Language Processing*, vol. 18, pp. 2067-2079, 2010.

[23] D. Povey, A. Ghoshal, G. Boulianne, L. Burget, O. Glembek, N. Goel, M. Hannemann, P. Motlíček, Y. Qian, P. Schwarz, J. Silovsky, G. Stemmer, and K. Veselý, "The Kaldi speech recognition toolkit," in *Proc. IEEE ASRU*, 2011.